# Young Massive Clusters/Associations in the GMC $G23.3-0.3$

M. Messineo[1,2,9], K. M. Menten[1], D. Figer[2], B. Davies[3], J. S. Clark[4], V. D. Ivanov[5], R.P. Kudritzki[6], R. Michael Rich [7], J. MacKenty[8], and C. Trombley[2]

**Abstract** An overview of a spectroscopic survey for massive stars in the direction of the Galactic giant molecular complex G23.3-0.3 is presented (Messineo et al. 2010, 2014). This region is interesting because it is rich in HII regions and supernova remnants (SNRs). A number of 38 early-type stars, a new luminous blue variable, and a dozen of red supergiants were detected. We identified the likely progenitors of the SNRs W41, G22.7-00.2, and G22.7583-0.4917.

## 1 Introduction

In the disk of the Milky Way massive stars continuously form. Their spatial distribution traces the disk morphology and the recent history of star formation. They chemically enrich the interstellar medium by losing mass at a high rate, and by exploding as SNRs. The modality of their formation is, however, still unclear. Massive stars are mostly found in massive stellar clusters ($> 10^4$ M$_\odot$), because there they are more abundant and easier to detect. Several massive stars are, however, born in isolation or ejected from the parent clusters (Oh, Kroupa & Banerjee 2014).

Mapping of massive stars in giant molecular clouds at different galacto-centric distances is a key ingredient for understanding the dependence of massive star formation on environments, in particular to understand starburst clusters and associations. We performed a search for evolved massive stars in the direction of the giant molecular complex $G23.3-0.3$.

## 2 Looking in the direction of the giant molecular complex $G23.3-0.3$

Several galactic structures intervene along the line of sight at a longitude of $23.3°$ and a latitude of $-0.3°$; one crosses the closest Sagittarius-Carina arm, and the Scutum-Crux arm, close to where the Scutum-Crux arm meets the Galactic bar. By integrating the $^{12}$CO data-cube over intervals of velocities, a giant molecular cloud appears in the 70-85 km s$^{-1}$ map (Dame et al. 1986; Albert et al. 2006). This cloud, hereafter named GMC G23.3−0.3, is associated with the supernova remnant W41 (SNR W41), and has an H$_2$ mass of at least $2 \times 10^6$ M$_\odot$ (Albert et al. 2006).

[1] Max-Planck-Institut für Radioastronomie, e-mail: messineo@mpifr-bonn.mpg.de
[2] Center for Detectors, Rochester Institute of Technology
[3] Astrophysics Research Institute, Liverpool John Moores University
[4] Department of Physics and Astronomy, The Open University
[5] European Southern Observatory
[6] Institute for Astronomy, University of Hawaii
[7] Department of Physics and Astronomy, University of California
[8] Space Telescope Science Institute
[9] European Space Agency (ESA)

As shown in Figure 1 and in Paper I (Messineo et al. 2010), there is an extraordinary spatial superposition of five supernova remnants onto the GMC G23.3−0.3. The CO intensity along the line-of-sight of four of these SNRs peaks at 70-85 km s$^{-1}$; the SNRs W41 (Green 1991), G22.7−0.2 (Green 1991), G22.7583−0.4917 and G22.9917−0.3583 are likely physically associated with the GMC G23.3−0.3 (Messineo et al. 2010). This GMC is at a kinematic distance of about 4-5 kpc (e.g. Albert et al. 2006).

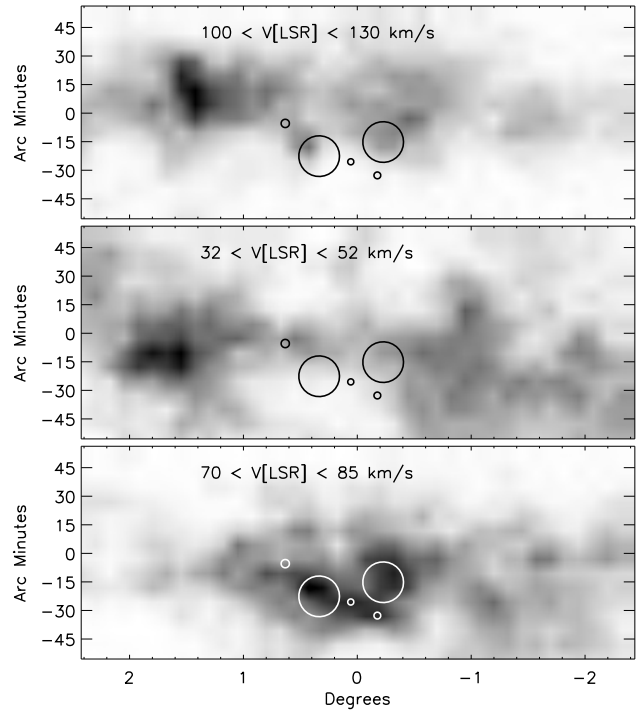

Fig. 1: CO maps centered at a longitude of 22.92° and a latitude of 0.07°; from the top to the bottom ranges of used velocities are 100-130 km s$^{-1}$, 32-52 km s$^{-1}$ and 70-85 km s$^{-1}$ respectively. The 70-85 km s$^{-1}$ is the GMC G23.3−0.3 (Albert et al. 2006; Messineo et al. 2010), which corresponds to object [23,78] in Dame et al. (1986). Superimposed are the locations of the supernova remnants W41 (Green 1991), G22.7−0.2 (Green 1991), G22.7583−0.4917 (Helfand et al. 2006), G22.9917−0.3583 (Helfand et al. 2006) and G23.5667−0.0333 (Helfand et al. 2006).

## 3 Massive stars and supernova progenitors.

In PaperII (Messineo et al. 2014) we performed a near-infrared spectroscopic survey of bright infrared stars in the direction of the GMC G23.3−0.3. We discovered 38 new early-type stars, of which 11 are O-type (see Figure 2), one is a new candidate Luminous Blue Variable and a few are red supergiants.

The O-type stars have interstellar extinction in $K_S$-band, $A_{K_S}$, from 1.34 to 1.90 mag, and likely belong to the GMC; several of the B-type stars are at a shorter distance ($A_{K_S} < 0.8$ mag). The massive O-type stars yield a spectrophotometric distance within errors consistent with that of the GMC. By assuming a distance of 5 kpc, we estimated for the O-types a range of masses from 25 to 60 $M_\odot$ and a likely age of 4-6 Myrs.

The detected O-type stars are not found in clusters, bur rather sparse on regions with radii of 7-8 pc (for a distance of 5 kpc). These sizes are similar to those of massive stellar associations (see Figure 3), such as Cyg OB2 (Wright et al. 2014). The core of W41 and the southern border of G22.7−0.2 are particularly rich in O-types; these two regions are separated by 0.5° (39 pc) but they share similar physical properties. This strengthens the physical associations of these population of massive stars with the SNRs, and a unique structure (GMC) at a distance of about 5 kpc. Massive stars were detected in the direction of the cores of three SNRs (O-types in two cores, red supergiants in one core). Beside distance, the association is based on morphology. In Figure 3, we overplot a radio image of the SNR G22.7583−0.4917 on a stellar infrared map and the O4 star #25 is located exactly on its center.

In the core of W41, which is populated by bizarre collapsed objects, such as magnetars and soft-gamma-ray-repeaters, the detection of massive stars allows placing contraints on the masses of their progenitors.

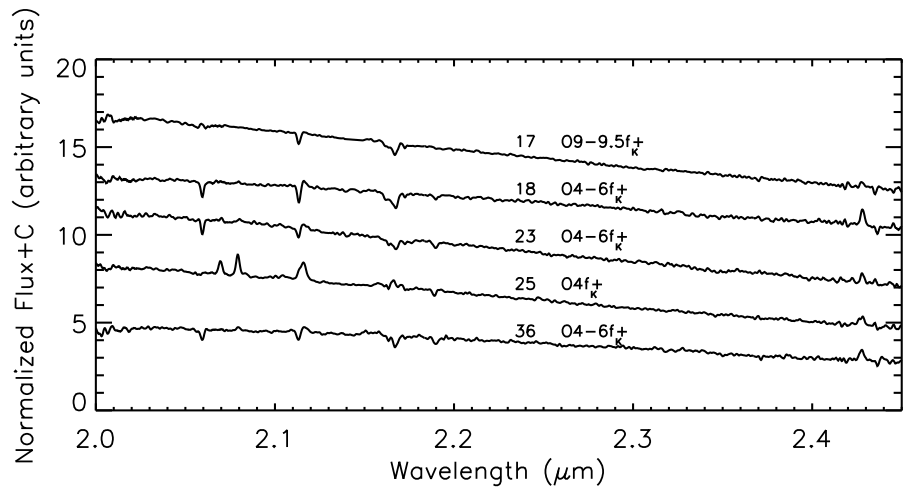

Fig. 2: A few normalized spectra of detected massive O-type stars. H I, He I, He II, N III, C IV, and Si IV lines were detected (Messineo et al. 2014).

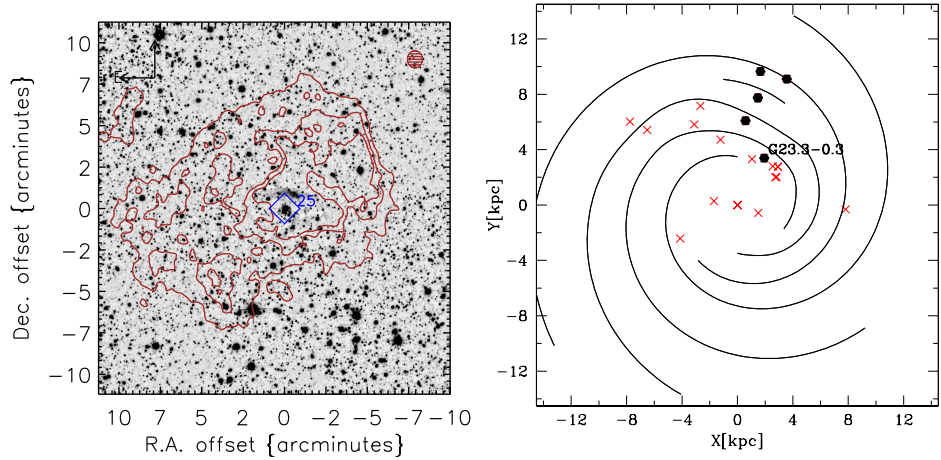

Fig. 3: Left side: Contours of the SNR G22.7583−0.4917 (Helfand et al. 2006), taken from the MAGPIS data at 20 cm (Helfand et al. 2006), are overlaid on a UKIDSS K-band image (Lucas et al. 2008). Contour levels are 3.0, 5.0 and 7.0 mJy beam$^{-1}$. The beam size is $6''.2 \times 5''.4$ FWHM and the position angle of the major axis is along the north-south direction. The massive O-type star #25, newly detected, is located at the center of the 20 cm emission (Paper II). Right side: Galactic distribution of known massive clusters and associations (within b=1°, and without central clusters) with masses $>\sim 10^4$ M$_\odot$. Spiral arms are from Cordes & Lazio (2002). The Galactic center is at (0,0) and the Sun is at (0,8). The location of "G23.3-0.3" is labeled. The position of the massive stellar clusters (crosses) and associations (hexagons) are taken from Figer et al. (2005); Pfalzner (2009); Messineo et al. (2008, 2009, 2011); Negueruela et al. (2010, 2011); Borissova et al. (2012); Davies et al. (2012a,b).

**Acknowledgements** MM thanks Prof. T. Dame for the CO data, and the conference organizers.